# On Optical Properties Of Ion Beams Extracted From Electron Cyclotron Resonance Ion Source

V. Mironov, S. Bogomolov, A. Bondarchenko, A. Efremov, and V. Loginov

Joint Institute for Nuclear Research, Flerov Laboratory of Nuclear Reactions, Dubna, Moscow Reg. 141980, Russia

**Abstract.** *Ion extraction from DECRIS-PM source is simulated by using initial distributions of ions at the extraction aperture obtained with NAM-ECRIS code. Three-dimensional calculations of plasma emissive surface are done and ions are traced in the extraction region. The ion beam profiles show strong aberrations due to shape of plasma meniscus; hollow beam features are reproduced, as well as changes in profiles for different focusing conditions.*

## 1. Introduction

Optical properties of the ions beams after their extraction from Electron Cyclotron Resonance Ion Source (ECRIS) strongly affect both ion losses during the beam transport in Low Energy Beam Transport Lines (LEBT) and efficiency of ion injection into accelerators. Emittances of the beams are mostly defined by the fact that the ions are extracted from the strong magnetic field of ECRIS toward the field-free drift space in LEBT, which causes the beam rotation with the final angles of ion propagation dependent on the ion initial radial position [1].

Apart from this factor, strong inhomogeneity of ions beam profiles is observed immediately after beam extraction that cannot be explained by aberrations in optical elements in LEBT. The profiles are hollow and triangular [2]; the beam divergence is different for the central and peripheral parts of the beam [3]. While the triangular shapes are easily explained by influence of the hexapolar component of magnetic field in ECRIS, the hollow profiles did not find clear explanation till now.

It is argued elsewhere that the ions are pushed out of the beam axis by strong radial electric field resulted from incomplete neutralization of the beam space charge [4]. Also, it is conjectured that the plasma shape inside the source is such that most of ions are born outside the source axis leading to the hollow beam profiles generation directly at the extraction aperture [5].

Hereby, we report on the results of dedicated numerical simulations of ion extraction from ECRIS, combining calculations of ion production in ECRIS by using NAM-ECRIS code [6,7,8] and three-dimensional calculations of plasma meniscus shape. Ions are injected from the meniscus into the beam line and traced until they reach the magnetic field-free region in front of the analyzing magnet, where the beam profiles are calculated for different strengths of the focusing solenoidal lens. The profiles are compared with the experimental results from [3] and good agreement is found between the calculations and experiments. Basing on the simulations, we come to conclusion that the observed peculiarities in profiles of ion beams are mainly determined by shape of the plasma meniscus.

The paper is organized in the following way: first, we give details of DECRIS-PM source design and present the NAM-ECRIS calculations that result in ion beam profiles at the extraction aperture; as the second step, we describe the extraction region of the source and 3-D calculations of the plasma meniscus shape; the final section of the paper deals with ion tracing in magnetic and electric field in the source extraction region and with discussion of the obtained beam parameters.

## 2. Ion production in DECRIS-PM

### A. The source parameters

The investigated source is the 14-GHz all-permanent magnet ECRIS designed for ion production for DC-280 cyclotron at FLNR JINR, Dubna [9]. The source is capable to produce up to (0.5-1) mA of $Ar^{8+}$ ions with around 500 W of injected microwave power, which makes it a good representative of its generation. The source chamber dimensions and magnetic field parameters are listed in Table 1.

Table 1. Parameters of DECRIS-PM source.

| Chamber diameter, cm | 7.0 |
|---|---|
| Chamber length, cm | 23.0 |
| Magnetic field at injection, T | 1.34 |
| Magnetic field at extraction, T | 1.11 |
| Magnetic field at minimum, T | 0.42 |
| Hexapole field at the wall, T | 1.1 |
| Extraction aperture, Ø, cm | 1.0 |
| Microwave frequency, GHz | 14.0 |

The magnetic field in the source is formed by four ring-like permanent magnets with varying directions of magnetization. To control the minimal magnetic field at the source, electrical coil is installed in between the magnets, which allow tuning the magnetic field profile in the range of ±0.075 T. Length of the resonance zone along the source axis is 9.5 cm and the zone is well centered inside the source chamber: distance between the $B_{min}$ position and the extraction plasma electrode is 5 cm. The hexapole magnetic structure consists of 24 segmented permanent magnets in Halbach configuration, the hexapole field at the radial walls is 1.1 T.

Plasma electrode is placed at local maximum of magnetic field at the extraction side of the source. Biased electrode with diameter of 3 cm is installed at the injection side of the source.

Ion extraction is done by using three-electrode structure, which consists of the plasma electrode, puller and the grounded electrode. Shape of electrodes is depicted in Fig.1, where electric potential distribution is shown as calculated with *POISSON* code [10].

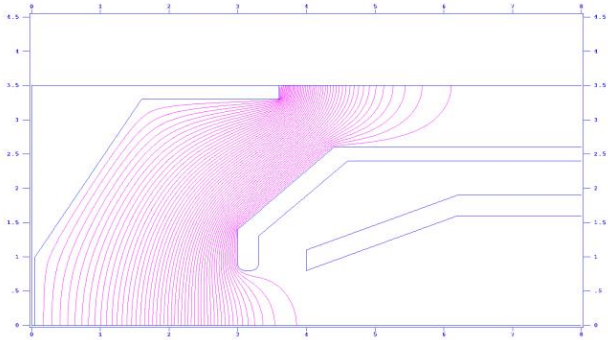

Fig.1 Extraction region of DECRIS-PM source.

The plasma electrode is angled by 46° in respect to the extraction axis; angling starts from radius of 1 cm; central part of the electrode is flat and has a thickness along z-axis of 0.5 mm. The electrode is flat inside the source. Extraction aperture has diameter of 1 cm. The puller is negatively biased up to -100 V and has the internal aperture of 1.6 cm and length of 10 cm; grounded electrode has the same opening of 1.6 cm and length of 12 cm. Angle of inclination for puller is 42° and 30° for the grounded electrode. Extraction gap is of 3 cm in the default configuration and can be adjusted when needed. Extraction voltage is up to 30 kV.

The specific feature of all-permanent magnet source is that the magnetic field outside the source expands for the relatively large distance; it changes its polarity at around 6 cm from the plasma electrode and reaches the level of 0.01 T at distance of 65 cm from the plasma electrode.

Focusing solenoid is placed at the extraction region to control the ion beam transport in LEBT. The solenoid length is 23 cm, including the yoke; the solenoid inner diameter is 11.2 cm. The solenoidal magnetic field is varied depending on the desired focusing strength.

   B. *Ion production*

The ion dynamics in the source is simulated by using *NAM-ECRIS* code, which is described in details elsewhere [6,7,8]. The code calculates ion production by using the Particle-in-Cell Monte-Carlo collision approach. Numerical particles are traced by taking into account thermalizing ion-ion collisions, charge-changing collisions with electrons and atoms, ion heating due to electron-ion collisions and incomplete energy absorption in collisions of particles with the walls. Ions are moving in the magnetic field of the source and in the electric field of the pre-sheath.

Ions are supposed to be confined within the ECR-limited volume by a small dip in positive plasma potential. For the electron energies used in the calculations and for 14 GHz microwave frequency, to define the ECR surface we select the resonance magnetic field value $B_{res}$=0.566 T.

The dip value is selected basing on dedicated simulations of electron dynamics with *NAM-ECRIS(e)* code [8] in assumption that the electric field magnitude at the resonance surface equals to 500 V/cm. For this specific study, the dip value is set to 0.02 V and the gas flow into the source is 0.7 pmA; this gas flow was selected such as to reproduce the experimentally measured extracted ion currents obtained with the source tuned to maximize the $Ar^{8+}$ current. We restrict ourselves to simulations of argon plasma.

We select the electron temperature inside the ECR volume equal to 45 keV as derived from the electron dynamics calculations, the electron temperature outside the volume ($T_{ec}$) is a free parameter fixed to 5 eV. The logics behind selecting this "cold" electron temperature is the following: we assume that the plasma potential drop in sheath $V_s$ is 20 V, close to the values measured by analyzing the energy spreads of the extracted ions [11] and with Langmuir probes [12], and we roughly estimate that $V_s \sim 4T_{ec}$. The cold electron temperature can vary in wide range, being dependent on the gas flow and microwave power injected into the source; also, there are indications that it can be controlled by the biased electrode. Indeed, reaction of the extracted ion currents to variations in the biased electrode voltage was shown [13] to be not connected to changes in the plasma parameters in the dense central parts of plasma, but to processes in the peripheral plasma and at the extraction regions. It is possible that the biased electrode influences the cold electrons by plugging the electrons by its negative voltage. Also, presheath voltage drop can be affected by the electrode.

When ions cross the ECR surface and leave the dense plasma region, they are supposed to be accelerated toward the walls by the presheath electric field. To calculate the field, we approximate the ECR volume by enclosing it with a cylindrically symmetric surface. Then two-dimensional electric field is calculated by using *POISSON* code with taking the surface-to-wall potential difference $V_{ps}$ as free parameter. We set the presheath potential drop equal to 2.5 V ($\sim 0.5 T_e$); it is checked that ion production processes as well the shape of ion spatial distributions on the plasma electrode are not influenced noticeably by varying this value in the range from 0.1 to 10 V. Energies of ions that approach the plasma electrode before entering the sheath follow the presheath voltage variations and scale as $Q \times V_{ps}$, where Q is the ion charge state.

For the selected input parameters, the globally defined ion and electron confinement times are 0.4 ms, the total extracted ion current is 3.3 mA, mean electron density inside the ECR volume is around $10^{12}$ cm$^{-3}$, and total power associated with flux of lost electrons to the source walls is ~200 W for the mean energy of lost electrons around 5 keV [8].

Spatial distribution of ions hitting the plasma electrode is shown in Fig.2. All ion positions are shown, without resolving their charge states.

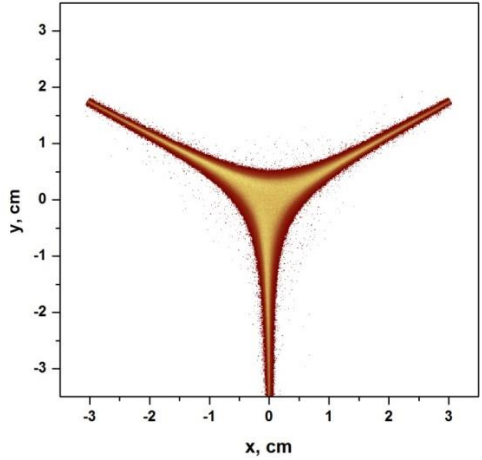

Fig.2 The calculated spatial distribution of ions at the plasma electrode

The lost ions form a triangular caused by superposition of solenoidal and hexapolar magnetic fields in the source. In experiments, these triangles are routinely seen as sputtered pattern at the plasma electrode. The calculations reproduce the narrow strips along the plasma star arms, and relatively weak halo around the arms. The distribution is peaked at the source axis, and those ions that leave the source through the extraction aperture form the extracted ion beams.

Parameters of the ions at the extraction aperture are taken as input for calculations of the ion extraction process. The charge-state-resolved parameters are listed in Table 2.

Table 2. Momenta of initial distributions of argon ions before their extraction

| Q | $\varepsilon_n^{rms}$, mm×mrad | $\langle r \rangle$, cm | $P$, mm×mrad | $E_{perp}$, eV | $E_{total}$, eV | $I_{extr}$, mA |
|---|---|---|---|---|---|---|
| 1 | 0.0122 | 0.344 | 0.051 | 0.29 | 2.64 | 0.07 |
| 2 | 0.0074 | 0.339 | 0.10 | 0.30 | 5.27 | 0.13 |
| 3 | 0.0071 | 0.333 | 0.15 | 0.31 | 7.78 | 0.19 |
| 4 | 0.0071 | 0.328 | 0.19 | 0.33 | 10.29 | 0.26 |
| 5 | 0.0075 | 0.324 | 0.23 | 0.36 | 12.78 | 0.33 |
| 6 | 0.0072 | 0.322 | 0.28 | 0.35 | 15.15 | 0.44 |
| 7 | 0.0077 | 0.319 | 0.32 | 0.39 | 17.5 | 0.46 |
| 8 | 0.0076 | 0.317 | 0.36 | 0.39 | 19.76 | 0.66 |
| 9 | 0.0078 | 0.316 | 0.398 | 0.41 | 22.12 | 0.42 |
| 10 | 0.0081 | 0.316 | 0.442 | 0.44 | 24.53 | 0.22 |

Here, the normalized rms-emittance is listed for different charge states of argon ions; the emittance is calculated, with using the standard notations, as

$$\varepsilon_n^{rms} = \frac{\beta}{2}[\langle x^2 + y^2 \rangle \langle x'^2 + y'^2 \rangle - \langle xx' + yy' \rangle^2]^{1/2} \qquad (1)$$

The emittance values are small and almost constant for all ion charge states. For singly charged argon ions, the emittance is larger than expected from global tendency due to non-negligible contribution of the ions that were created in the presheath by ionizing collisions of atoms with electrons and in ion-atom charge-exchange reactions.

The normalized measure of canonical angular momentum $P$ is calculated [14] as

$$P = \frac{\langle P_\Theta \rangle}{2mc} = \beta \left[ \langle xy' \rangle - \langle yx' \rangle + Q \frac{B}{2mc} \langle x^2 + y^2 \rangle \right] \qquad (2)$$

The mechanical term of the canonical angular momentum, $\beta[\langle xy' \rangle - \langle yx' \rangle]$, is negligibly small for the analyzed distributions, but will increase after extraction in the decreasing magnetic field following conservation of the canonical angular momentum, such that $\varepsilon_n^{rms}$ will be close to $P/2$ values.

The mean radius of the ion spatial distribution $\langle r \rangle = \sqrt{\langle x^2 + y^2 \rangle}$ is smaller for the high charge states than the value for the uniform distribution of 0.33 cm, and decreases slowly with Q, indicating weak accumulation of highest charge state ions toward the source axis. Influence of this accumulation does not change the global tendency of linear increase of the canonical momentum with the ion charge state.

Perpendicular energies of the ions $E_{perp}$ are relatively small and increase with increasing Q from 0.3 (Q=1) to 0.44 eV (Q=10). The energies are defined mostly by temperature of the ions confined inside the dense parts of the plasma by potential dip and by scattering of ions during their transport in presheath in collisions with cold electrons and in ion-ion collisions. The total energy of ions $E_{tot}$ scales linearly with the ion charge state, being mainly defined by the presheath potential drop. We notice here that for the highest charge states of argon ions (Q≥8) the energy is lower than the presheath value, while for the lowest charge states it is higher. This is caused by ion friction in presheath because of the ion-ion collisions, when the more energetic ions are slowed down and lower charge states are accelerated. There are both experimental and theoretical indications that the ion-ion interaction in presheath can be much stronger than predicted by classical ion-ion collision rate due to instability-enhanced ion friction [15]. In this case, energy spreads of the extracted ions will be larger than listed in Table 2, with the resulting increase in the initial emittances. We leave at the moment these issues open for the further investigation.

In the last column of Table 2, extracted ion currents are shown indicating that up to 0.65 mA of $Ar^{8+}$ current can be extracted in the given plasma conditions. The charge state distribution is close to what is measured for DECRIS-PM source at its optimal conditions.

From the ion distributions it is straightforward to calculate the electron density spatial distribution by observing that local current density is $j = Qen_i v_i$, where $v_i$ is the ion velocity and $n_i$ is the ion density. Ion densities are summed up and electron density is set equal to this sum from requirement of charge neutrality in plasma in front of extraction aperture. The resulting electron density transversal distribution inside the extraction aperture is shown in Fig.3 for x and y coordinates in the range of (-0.5÷0.5) cm on a mesh with step of 0.02 cm.

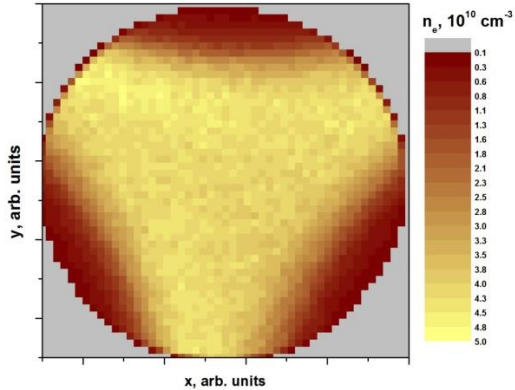

Fig.3 Electron density distribution at extraction aperture.

The distribution is a slightly hollow triangle. Maximal electron density is $5\times 10^{10}$ cm$^{-3}$, and the density decreases in between the plasma arms at the extraction aperture boundary to as low as $10^9$ cm$^{-3}$, more than by order of magnitude. Along the arms, density varies by ~ (10-15) % with local maxima close to the aperture boundary. The electron density is defined with an accuracy defined by the presheath voltage and scales as $V_{ps}^{-1/2}$.

### 3. Plasma meniscus

Ion extraction from the plasma proceeds along normal to boundary between plasma and vacuum in the extraction gap. Ions are accelerated in a thin sheath of a few Debye lengths acquiring there the energy of $Q\times V_s$. The boundary of plasma emitter or meniscus is free-moving in a sense that it is not in contact with any solid electrode.

In our calculations, for defining the meniscus shape we adopt the approach of Astrelin et al. [16]. It is argued there that the boundary position can be calculated from requirement

$$n_e T_e = \frac{\varepsilon_0 E_B^2}{8\pi} \qquad (3)$$

where $n_e$ and $T_e$ are the electron density and temperature respectively, and $E_B$ is the external electric field along the magnetic field at the point where the surface is calculated. Meaning of the requirement is that the electron pressure is balanced by electric field tension at each point of steady plasma emissive boundary. For electron temperature of 5 eV and electron density of $10^{11}$ cm$^{-3}$, the balancing electric field from Eq.3 is 4.77 kV/cm.

To find the meniscus shape for given electron density distribution, we use FORTRAN library from [17]. It applies boundary element method for solving 3D Laplace problem inside the domain interior to a closed surface. The extraction region of DECRIS-PM is approximated with a surface that consists of planar triangular elements, with around 300 mesh vertexes inside the extraction aperture. To improve numerical accuracy, we construct the mesh with closing it by the equipotential surface V=+3 kV adjacent to the puller electrode, keeping the plasma electrode and plasma boundary at potential of +20 kV. Plasma boundary is taken as an equipotential surface, neglecting the positive plasma potential value in respect to the plasma electrode and possible variations of the plasma potential in transverse directions. From this assumption, it follows that electric field at the plasma surface is directed perpendicular to it.

At each step of calculations, electric field component $E_z$ is calculated close to vertexes at extraction aperture (we assume that the magnetic field has only z-component there, such that $E_B=E_z$). The field is compared to the prescribed value from Eq.3, and vertexes are moved in z-direction along the source axis in iterative way to minimize the difference between the values. Typically, it requires around 250 iterations to achieve the solution converged such that the required and calculated fields differ by factor less than $10^{-3}$. Relatively slow solver of [17] defines the computational time of around 10 hours per run.

First, we calculate the meniscus shape for uniform distribution of electron density in the extraction aperture equal to $5 \cdot 10^{10}$ cm$^{-3}$. The shape is shown in Fig.4 with original dimensions and after multiplying z-coordinates by factor of 10 to visualize the shape in more details. Plasma is limited by extraction aperture of 1-cm diameter; flat part of the plasma electrode with outer diameter of 2 cm around the meniscus is also shown in Fig.4. Plasma expands into the extraction gap to the right, and the outermost z-position is 0.55 mm from the extraction electrode surface. Meniscus shape is close to flat for given geometry of extraction electrode and puller. Irregularities are seen at transition from plasma to the extraction electrode, where solution has a rather poor quality.

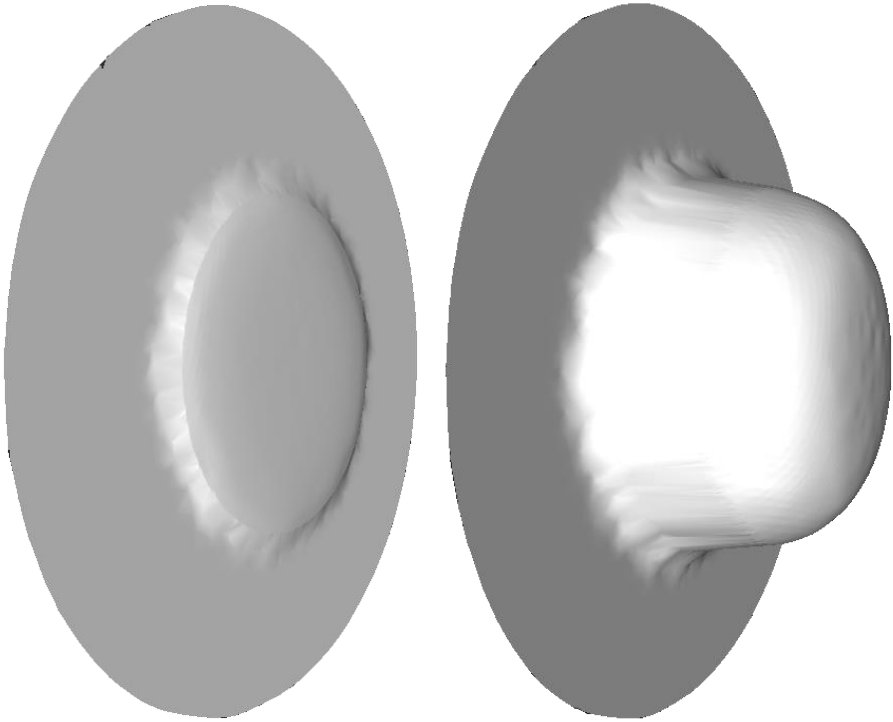

Fig.4 Meniscus shape for uniform electron pressure distribution with $n_e=5\cdot10^{10}$ cm$^{-3}$ and $T_e=5$ eV: a) original size, b) after multiplying z-coordinates by factor of 10.

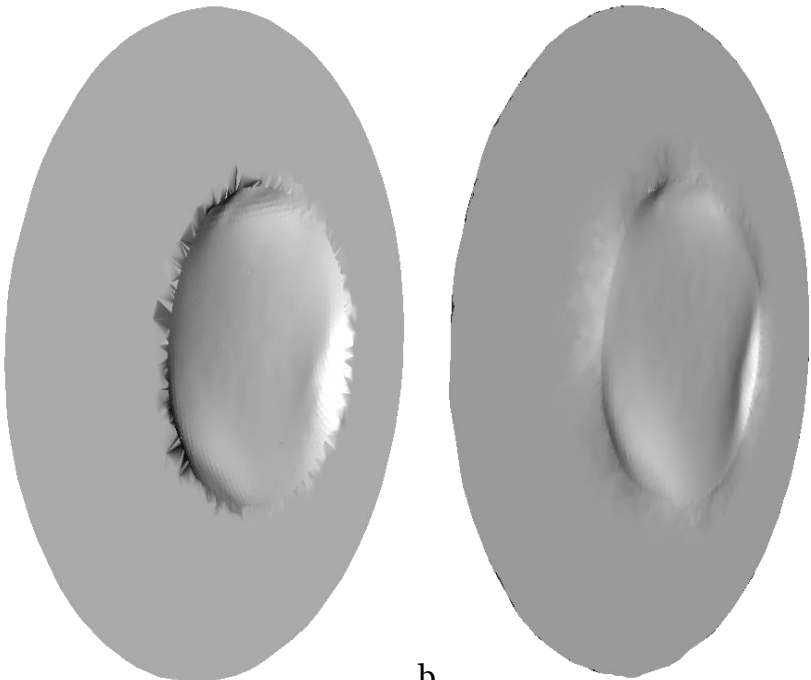

a                              b

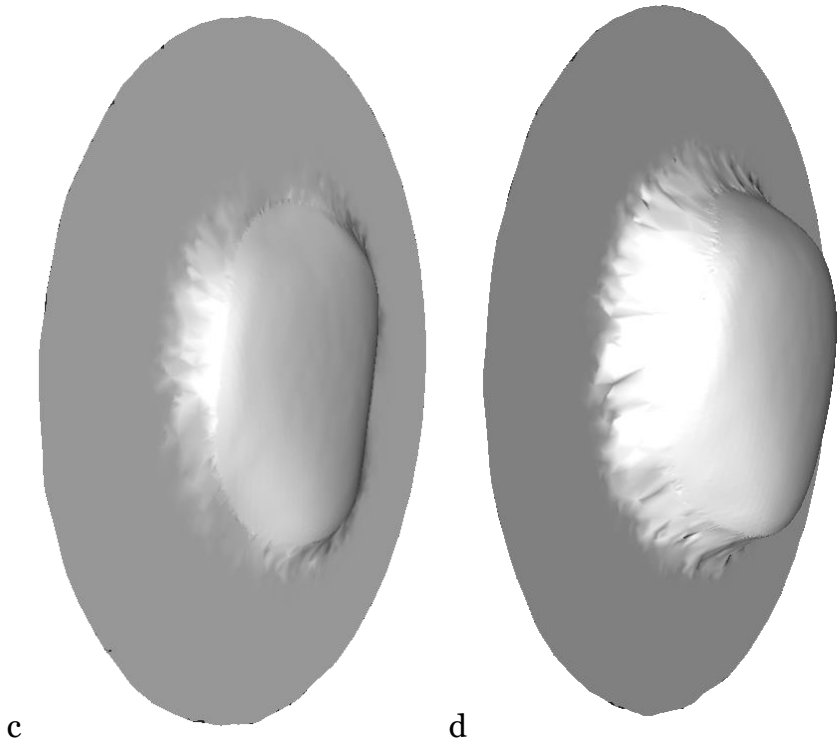

c                      d

Fig.5 Meniscus shapes for scaling the electron pressure distribution from NAM-ECRIS code by factor S of 0.5 (a), 1.0 (b), 2.0 (c) and 4.0 (d)

In Fig.5, we show the calculated meniscus shapes using the electrond density distribution as obtained in previous section, for different scalings of plasma pressure, namely after multiplying all values of electron density by factor S of a) 0.5, b) 1.0, c) 2.0 and d) 4.0. The larger is the electron pressure, the further downsteam is the plasma boundary; variations in electron density manifest themselves in variations of plasma boundary position and angle of inclination of the boundary in respect to the source axis. For scaling of the pressure equal to 1, variations in z-positions of the plasma boundary are from 0.18 mm for most dense parts to -0.34 mm for the least dense parts of plasma. Component of the vector normal to the plasma boundary along z-axis ($n_z$) varies from 1 to 0.75 for these settings.

Details of plasma boundary for S=1 are shown in Fig.6, where the boundary is shown after multiplying the longitudinal coordinates z by factor of 10. Triangular type of the boundary is more visible there, as well as variations of meniscus angle of inclination. Plasma density is coded in Fig.6 by colors, with the large density depicted in red and the low density in blue.

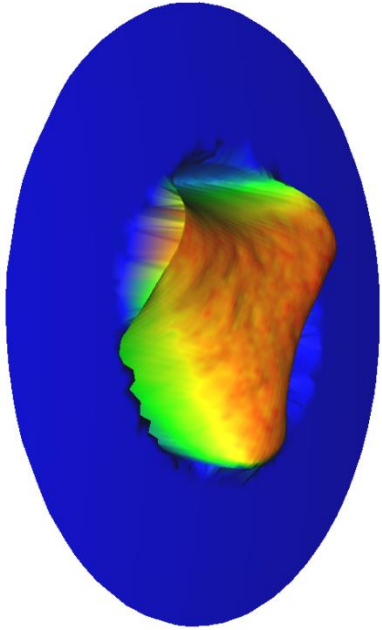

Fig.6 Meniscus for S=1 after multiplying all z-coordinates by factor of 10.

After calculations of the plasma emissive boundary, the mesh is imported into COMSOL Multiphysics 3D electrostatics module [18] and final calculations of electric fields in extraction region are done there. The results are imported as array of interpolated values ready to be used in our particle-tracing code. The typical output of COMSOL for z-component of electric field ($E_z$) close to the plasma electrode is shown in Fig.7. Here, the field is plotted at z=1 mm from the plasma electrode and for the scaling of electron pressure S=1.

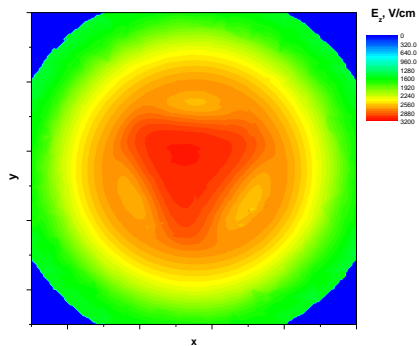

Fig.7 Electric field component in z direction ($E_z$) close to the plasma emissive boundary (z=0.1 cm).

Blue color in the plot indicates regions beyond the inclined plasma electrode surface; regions close to the extraction aperture is seen as yellow-red area. Electric field is largest at the source axis and along the plasma star arms (denser parts of the plasma) because they are closer to the measurement plane. We see strong perturbation of electric field by plasma boundary, which influences the extraction dynamics of ions in this region.

## 4. Ion tracing in extraction region

To calculate ion extraction dynamics in extraction region of DECRIS-PM source, we wrote special ion tracing code that follows the ion movement in external magnetic and electric fields. Self-induced magnetic and electric fields are not included, calculations are done without taking into account the space-charge effects. Initial positions and velocities of ions are taken from arrays prepared in *NAM-ECRIS* code. Velocities are incremented along normal to the plasma emission surface such as to take into account ion acceleration in the sheath, $\Delta E = Q \times V_s$, where $V_s = 20$ V. Typically, movement of $2 \cdot 10^5$ particles for each charge state is calculated. Profiles and emittances of ions are obtained at magnetic-field free position far away from the extraction electrode (z=88 cm) prior the ion entrance into the bending magnet. Focusing solenoid strength is selected to ensure the beam focusing at this position for specific charge states of ions.

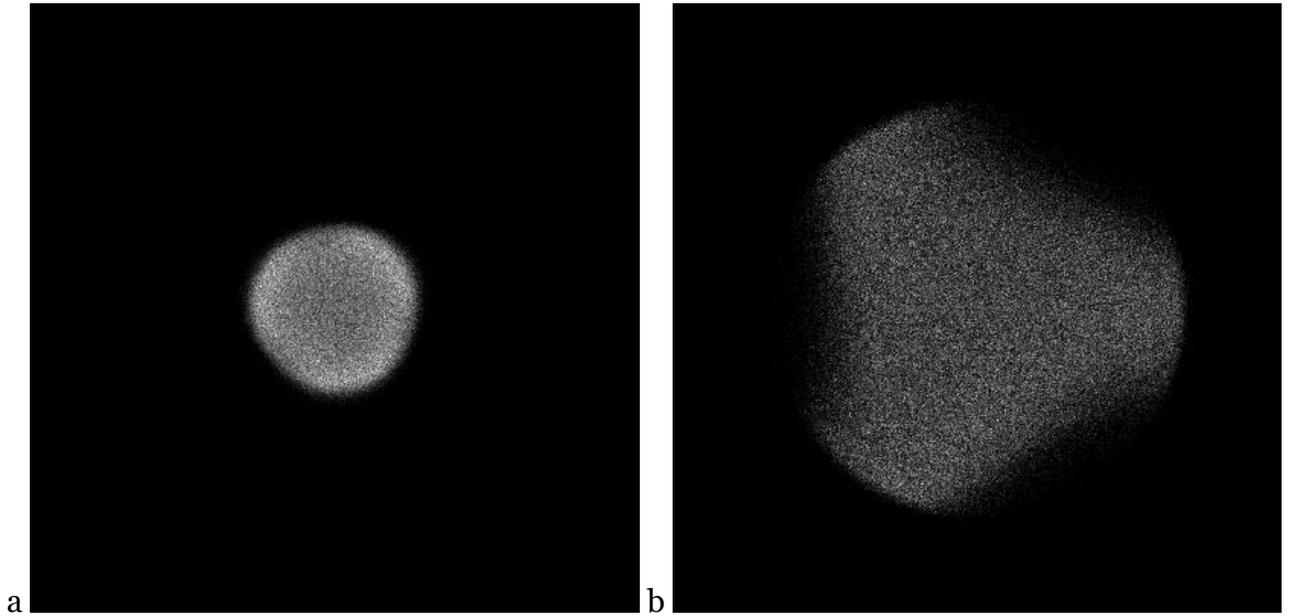

Fig.8. Profiles of $Ar^{1+}$ (a) and $Ar^{8+}$ (b) ions after their extraction at z=88 cm for flat meniscus.

We begin with calculations for flat meniscus with placing the plasma emissive surface at z=0 to check aberrations and general dynamics in this idealized situation. Ions are launched according to their distributions calculated by NAM-ECRIS and are accelerated in the sheath along z-axis. Profiles of $Ar^{1+}$ and $Ar^{8+}$ ion beams are shown in Fig.8. Triangular shape of the ion beam is visible, especially for the highly charged $Ar^{8+}$ ions, the profiles are rotated by magnetic fields of the source and of the focusing solenoid. No strong aberrations are seen; profile of $Ar^{1+}$ ions is slightly hollow reflecting its initial spatial distribution. Normalized emittances of the extracted beams are listed in Table 3 in [mm·mrad] units.

Two values are listed, the emittance as calculated according to Eq.1 and the emittance "cleaned" from the mechanical momentum as

$$\varepsilon_{n0}^{rms} = \frac{\beta}{2}[\langle x^2 + y^2\rangle\langle x'^2 + y'^2\rangle - \langle xx' + yy'\rangle^2 - \langle xy' - yx'\rangle^2]^{1/2} \quad (4)$$

The "full" emittance values for flat meniscus and for electron density from *NAM-ECRIS* are close to what is predicted by calculations of the canonical angular momenta (=0.5P) listed in Table 2. The "cleaned" values are close to the initial emittances indicating small aberrations in the extraction region for the flat plasma boundary and overall accuracy of field interpolations and particle tracing procedure.

Momenta the beam distribution calculated in assumption of uniform plasma pressure profile at the extraction aperture were obtained by uniformly launching the test particles from the meniscus shown in Fig.4, previous section. The rms-emittance values are larger than for the case of calculations with pressure profiles from *NAM-ECRIS* due to larger initial size. The "cleaned" emittances are close to the ones from flat meniscus

For simulations with taking into account the meniscus shape and pressure profiles from *NAM-ECRIS*, strong aberrations are seen in extracted ion beams. The corresponding emittance values are listed in Table 3 for scaling the electron pressure values with the factors 0.5, 0.75, 1.0, 2.0 and 4.0. Aberrations are minimized at the scaling S=1.0. Transmission factor for ions reaching the detection plane are close to 1 for all scaling factors except the extreme case of S=4.0, where losses can be as high as 33% for singly charged argon ions (the transmission factor is shown for this case in the separate column).

Table 3. Emittances of the extracted argon ions for the flat meniscus, for uniform electron pressure and for meniscus with the electron pressure scaled by factor S.

| | flat | | uniform | | S=1 | | S=0.5 | |
|---|---|---|---|---|---|---|---|---|
| Q | $\varepsilon_n^{rms}$ | $\varepsilon_{no}^{rms}$ | $\varepsilon_n^{rms}$ | $\varepsilon_{no}^{rms}$ | $\varepsilon_n^{rms}$ | $\varepsilon_{no}^{rms}$ | $\varepsilon_n^{rms}$ | $\varepsilon_{no}^{rms}$ |
| 1 | 0.027 | 0.007 | 0.029 | 0.008 | 0.033 | 0.020 | 0.039 | 0.03 |
| 2 | 0.052 | 0.0077 | 0.057 | 0.0097 | 0.058 | 0.027 | 0.065 | 0.04 |
| 3 | 0.075 | 0.0083 | 0.085 | 0.011 | 0.08 | 0.029 | 0.086 | 0.046 |
| 4 | 0.097 | 0.0088 | 0.11 | 0.012 | 0.10 | 0.031 | 0.11 | 0.047 |
| 5 | 0.12 | 0.0095 | 0.14 | 0.014 | 0.12 | 0.033 | 0.13 | 0.047 |
| 6 | 0.14 | 0.0095 | 0.17 | 0.015 | 0.14 | 0.034 | 0.15 | 0.045 |
| 7 | 0.16 | 0.0094 | 0.2 | 0.016 | 0.16 | 0.035 | 0.17 | 0.043 |
| 8 | 0.18 | 0.0090 | 0.22 | 0.017 | 0.18 | 0.036 | 0.19 | 0.041 |

| | S=0.75 | | S=2.0 | | S=4.0 | | |
|---|---|---|---|---|---|---|---|
| Q | $\varepsilon_n^{rms}$ | $\varepsilon_{no}^{rms}$ | $\varepsilon_n^{rms}$ | $\varepsilon_{no}^{rms}$ | $\varepsilon_n^{rms}$ | $\varepsilon_{no}^{rms}$ | $f$ |
| 1 | 0.036 | 0.025 | 0.034 | 0.022 | 0.038 | 0.033 | 0.68 |
| 2 | 0.061 | 0.033 | 0.059 | 0.029 | 0.053 | 0.037 | 0.75 |
| 3 | 0.082 | 0.035 | 0.082 | 0.034 | 0.071 | 0.042 | 0.81 |
| 4 | 0.10 | 0.036 | 0.10 | 0.038 | 0.093 | 0.051 | 0.86 |
| 5 | 0.12 | 0.037 | 0.12 | 0.041 | 0.12 | 0.064 | 0.91 |
| 6 | 0.14 | 0.037 | 0.15 | 0.044 | 0.15 | 0.074 | 0.95 |
| 7 | 0.17 | 0.038 | 0.17 | 0.047 | 0.17 | 0.083 | 0.97 |
| 8 | 0.18 | 0.038 | 0.19 | 0.049 | 0.20 | 0.090 | 0.99 |

Total emittance values are defined by the mechanical term due to the beam rotation in decreasing magnetic field, except the case of S=4, where the beam losses are so high that emittances become be lower of the expected values for some charge states.

Data of Table 3 are presented in graphical form in Figs.9 and 10. In Fig.9, the "cleaned" emittance is shown for different scaling factors of electron pressure distributions, and for the flat meniscus and for uniform distribution of electron pressure. In Fig.10, the ratio between the normalized emittance and the value predicted by conservation of the canonical angular momentum, minus 1, is plotted.

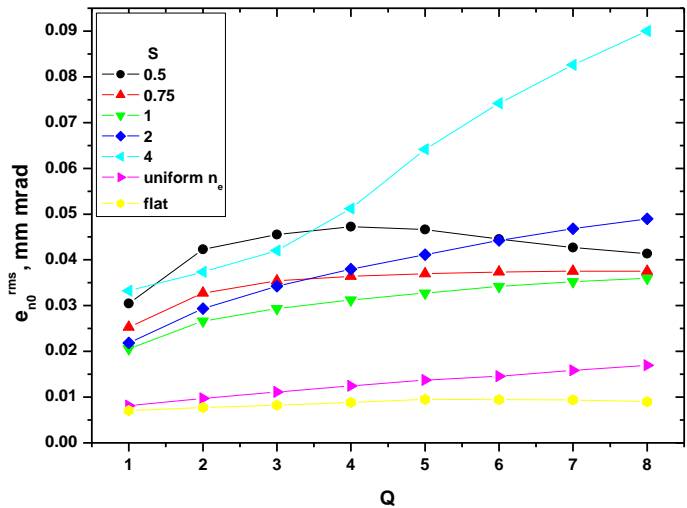

Fig.9 "Cleaned" normalized emittance for different scaling factors of electron pressure and for the flat meniscus.

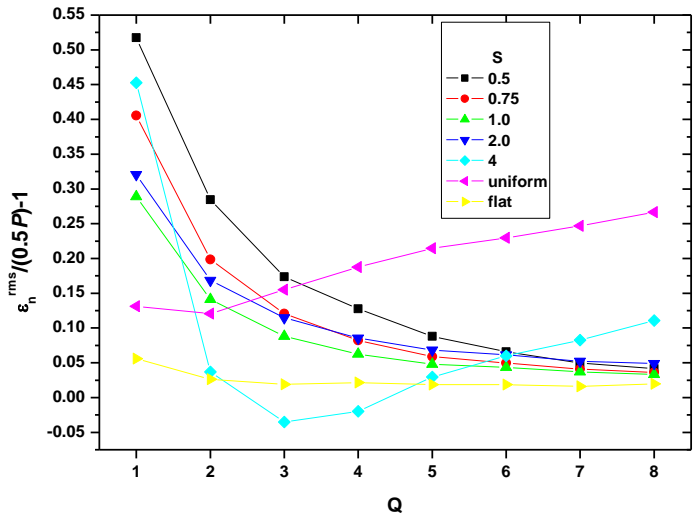

Fig.10 Ratio of the normalized emittance to the value defined by the canonical angular momentum conservation ($P/2$).

Aberrations of the beams due to non-uniform meniscus shape can be better understood by observing the beam focusing with varied focusing solenoid strength. For the case of S=1 and for the $Ar^{1+}$ ions, profiles of ion beam at the detection plane are shown in Fig.11. Profiles are shown in the box with dimensions of (-1÷+1) cm in x and y directions.

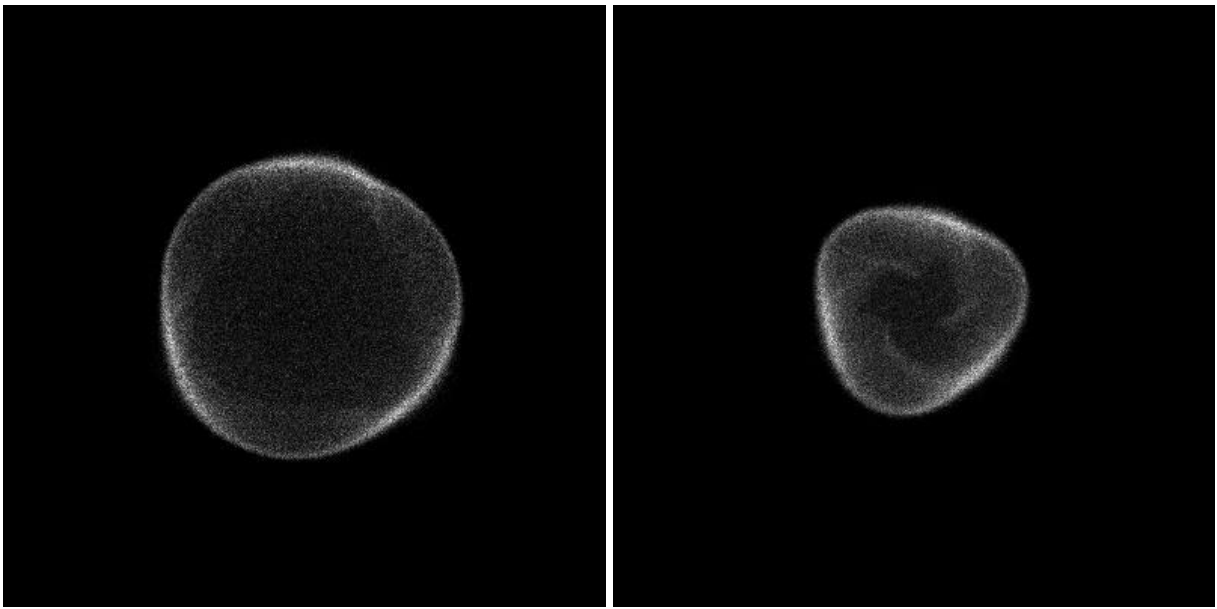

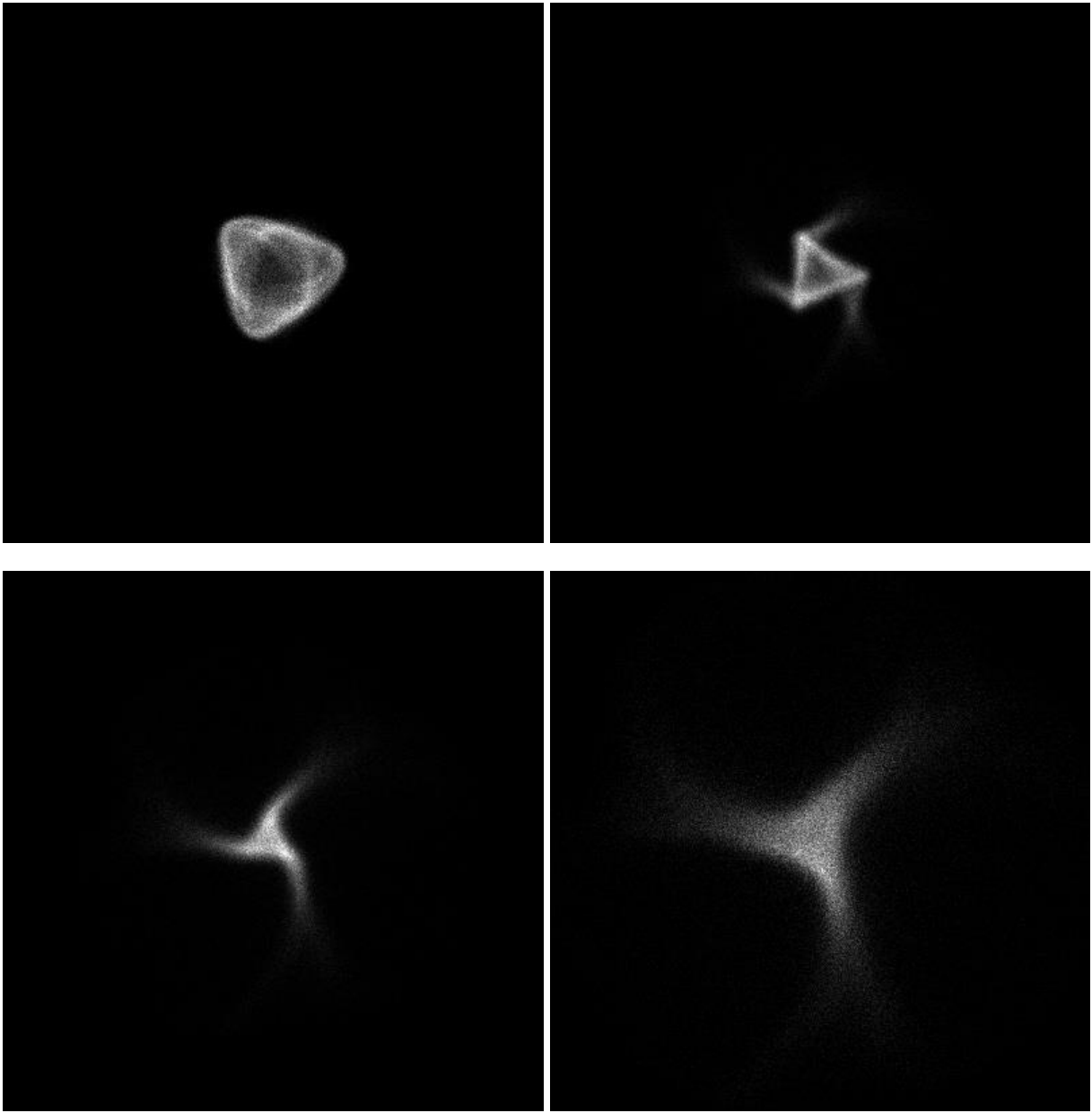

Fig.11 Profiles of $Ar^{1+}$ ion beam at z=88 cm for different focusing solenoid fields.

Basically the same beam profiles are observed for other charge states of ions. The profiles are hollow structures with stronger focusing of those parts of the beam that lay along the initial plasma star arms. After reaching the sharpest focus, triangular shape is emerging with distortions due to the beam rotation in magnetic field.

Dynamics of the beam profiles is very close to the experimentally observed features [3]. We connect the calculated aberrations of the extracted ion beams to variations in the angles of ion propagation close to the plasma boundary. In "object-image" terminology of geometrical optics, plasma meniscus shape changes the object positions for different parts of the beam.

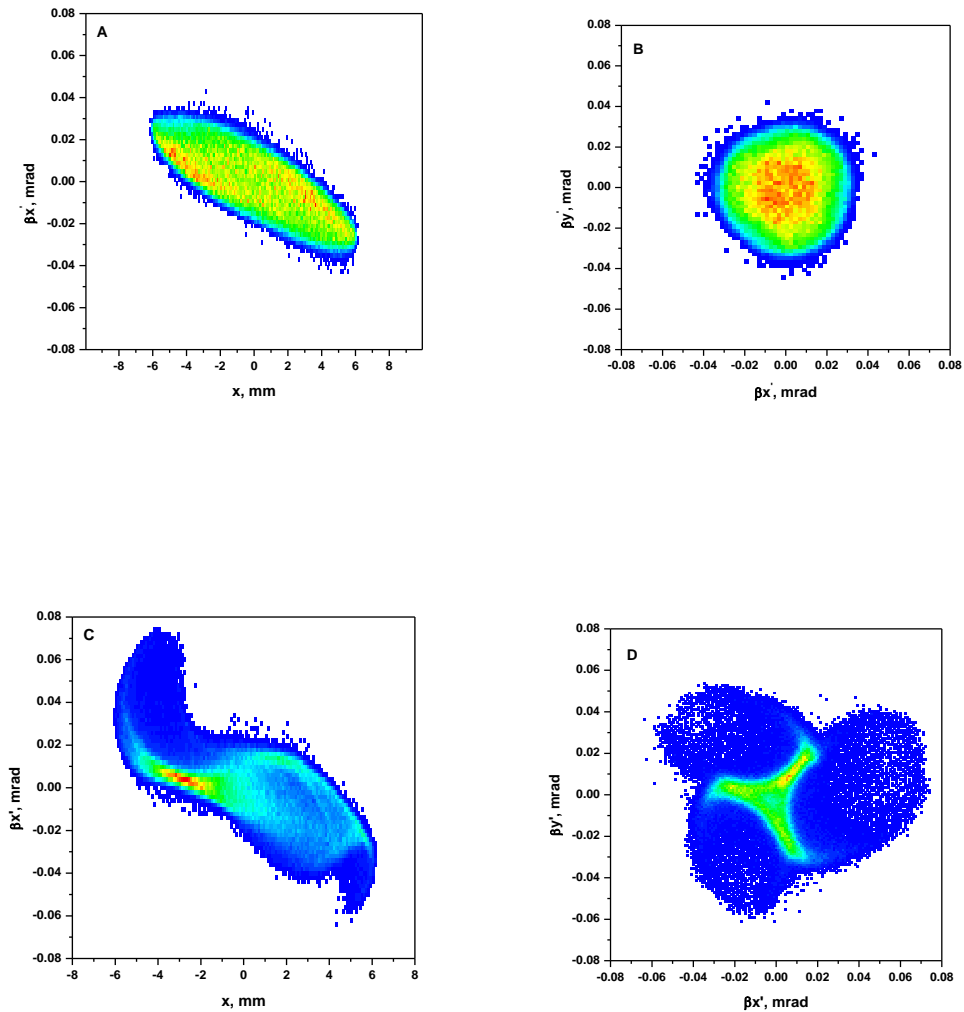

Fig.12 Phase portraits of Ar$^{1+}$ ion beam with flat meniscus (a -b) and for S=1 case (c-d).

In Fig.12, emittance plots are shown for x-x' plane and correlation plot for x'-y' projection with using as the starting conditions the flat meniscus (a-b) and the meniscus for scaling factor S=1 of the electron pressure (c-d). Beam of Ar$^{1+}$ ions is presented for the same focusing solenoid magnetic field as the first profile in Fig.11. Strong aberration is seen in x-x' emittance for S=1 case, as well as correlation between x' and y' components that shows the initial triangular feature of the meniscus shape.

Particle density at the (x'-y') plots with flat and distorted meniscuses is essentially the same at its maxima, indicating that the correlated pattern in Fig.12-C emerges not from concentration of ions by some reasons into triangle, but due to broader distribution of particles in the phase space outside the structure. Triangles of Fig.11 are formed by strong aberrations in areas beyond the visible pattern.

Concluding, we observe that plasma meniscus shape strongly influences the extracted beam profiles. The calculated beam profiles are close to what is measured in experiments, including the hollow beam formation and dynamics of the profiles with different focusing in beam line. The revealed influence of the meniscus allows explaining the previously reported changes in the beam profiles for different microwave frequencies in ECRIS [19], as well as changes in the beam transmission efficiency for optimized extraction electrode geometry [20].

Uniform distribution of plasma density across the extraction aperture gives the lowest distortions of the beam profiles, suggesting the way to improve the beam quality by decreasing the extraction magnetic field in ECRIS. This is, however, accompanied with the corresponding decrease in the extracted ion currents and increase in the total emittance due to larger initial size of the beam, and compromise should be found by optimizing the field.

Best quality of the beam is observed when the meniscus is close to flat. This is an open question whether the biased electrode influences the plasma density close to the extraction and allows optimization of meniscus shape for specific plasma conditions. Implicit support of this scenario stems from our experimental observations of dependence of the optimal biased electrode voltage on the extraction voltage ($V_{extr}$) at DECRIS-PM source. With increasing the extraction voltage, it is needed to increase the negative voltage of the biased electrode to maximize the extracted currents of highly charged argon ions; the optimal voltage increases from (-100 V) at $V_{extr}$=10 kV to approximately (-300 V) for $V_{extr}$=20 kV.

For all investigated cases, emittances of the extracted beams are close or exceed the values defined by conservation of canonical mechanical momentum, except the case of S=4, where losses of ions are so large that emittance is lower of the expected value. There are measurements of extracted ion beam parameters [21] that indicate the emittances lower of the "magnetic emittance" limit by factor of ~2. It is conjectured that such behavior can be explained by concentration of highly charged ions at the source axis and smaller sizes of their initial distributions. We see the ion concentration on the axis, but not so strong as to explain the observations. It would be interesting to estimate possible losses of ions during their extraction in experiments like [21].

Apparently, optimized electrode geometry should exist that minimizes the beam distortions and it is planned to continue the investigations for different electrode shapes. Tracing of the beams with the obtained profiles through the bending magnet and other optical elements in beam-line is also planned in the future.